\documentclass[useAMS,usenatbib]{mn2e}

\usepackage{graphicx}
\bibpunct{(}{)}{;}{a}{}{,}
\usepackage{txfonts}

\title[The weak magnetic field of PDS\,2]{
The prevalence of weak magnetic fields in Herbig~Ae stars: \\ The case of PDS\,2
}

\author[Hubrig et al.\ 2015]{
S.~Hubrig$^1$\thanks{E-mail: shubrig@aip.de},
T.~A.~Carroll$^1$, M.~Sch\"oller$^2$, I.~Ilyin$^1$ \\
$^1$ Leibniz-Institut f\"ur Astrophysik Potsdam (AIP), An der Sternwarte~16, 14482~Potsdam, Germany\\
$^2$ European Southern Observatory, Karl-Schwarzschild-Str.~2, 85748~Garching, Germany
}

\begin{document}

\date{Accepted Received; in original form}

\pagerange{\pageref{firstpage}--\pageref{lastpage}} \pubyear{2015}

\maketitle

\label{firstpage}

\begin{abstract}
Models of magnetically driven accretion and outflows
reproduce many observational properties of T\,Tauri stars, 
but the picture is much less clear for the Herbig~Ae/Be stars,
due to the poor knowledge of their magnetic field strength and topology.
The Herbig~Ae star PDS\,2 was previously included in two magnetic studies 
based on low-resolution spectropolarimetric observations. 
Only in one of these studies the presence of a weak mean longitudinal
magnetic field was reported.
In the present study, for the first time, 
high-resolution HARPS spectropolarimetric observations of PDS\,2
are used to investigate the presence of a magnetic field. 
A firm detection of a weak longitudinal magnetic field is achieved using the multi-line 
singular value decomposition method 
for Stokes profile reconstruction ($\left<B_{\rm z}\right>=33\pm5$\,G). 
To gain better knowledge of typical magnetic field strengths in late Herbig Be and Herbig Ae stars,
we compiled previous magnetic field measurements, revealing that only very few stars have 
fields stronger than 200\,G, and half of the sample possesses fields
of about 100\,G and less.
These results challenge our current understanding of the magnetospheric accretion in intermediate-mass pre-main 
sequence stars as they indicate that 
the magnetic fields in Herbig~Ae/Be stars are by far weaker than those measured in T\,Tauri stars.
\end{abstract}


\section{Introduction}

Recent observations of the disk properties of intermediate mass 
Herbig~Ae and late Herbig~Be  stars suggest a close parallel to T\,Tauri stars, revealing the same size range of 
disks, similar optical 
surface brightness and similar structure consisting of inner dark disk and a bright ring.
Models of magnetically driven accretion and outflows
successfully reproduce many observational properties of low-mass
pre-main sequence stars, the classical T\,Tauri stars. However, due to the very poor knowledge of the magnetic field 
strength and magnetic field topology in Herbig stars, current theories are not able to present a 
consistent scenario of how the magnetic fields in Herbig~Ae/Be stars are generated and how these fields interact 
with the circumstellar environment, consisting of a combination of accretion disk, magnetosphere,
and disk-wind region or jets.

As of today, only about 20 late Herbig~Be and Herbig~Ae stars have been reported to possess large-scale organized 
magnetic fields 
(e.g., \citealt{Hubrig2009}; \citealt{alecian2013a}), 
using low- and/or high-resolution spectropolarimetric observations.
Moreover, only for the two Herbig~Ae stars HD\,101412 and V380\,Ori 
\citep{Hubrig2011a,alecian2009},
the magnetic field geometry has been constrained in previous studies. It is very likely that 
the rather low detection rate
of magnetic fields in Herbig~Ae stars can be explained by the weakness of these fields and/or 
by rather large measurement uncertainties. Indeed, in the currently largest high-resolution spectropolarimetric survey 
of the magnetic field in these stars by \citet{alecian2013a} with 132 measurements for 70 Herbig stars, 
the measurement uncertainty is worse than 200\,G for 35\% of the measurements, and 
for 32\% of the measurements it is between 100 and 200\,G, i.e.\ only 33\% of the measurements showed a
measurement accuracy below 100\,G.
Therefore, any new detection/confirmation of the presence of a magnetic field in a Herbig~Ae/Be star
is important to increase the sample of magnetic Herbig stars. Clearly, only after we gain knowledge of 
the magnetic field strength and its topology, can we start to understand how it interacts with the 
circumstellar environment.

Polarimetric observations of the Herbig~Ae star PDS\,2 (CD\,$-$53$^{\circ}$\,251) 
were obtained on two different epochs using the HARPS polarimeter.
This star has been identified as a Herbig~Ae candidate star in the 
Pico dos Dias Survey by  \citet{greg1992}. PDS\,2 displays H$\alpha$ in emission, and 
the pre-main sequence nature is indicated by its {\it IRAS} colours.
The mass accretion rate was determined by \citet{pog2012}
on eight epochs using different spectral accretion indicators from the near-UV to the near-IR.
This work revealed short-term night-to-night variability, as well as  
long-term variability on a time scale of tens of days.    

Using the 0.6\,m R.E.M. telescope on La~Silla, a preliminary analysis of PDS\,2 by \citet{Berna2007} 
showed the presence of $\delta$~Scuti-like pulsations with  three frequencies
on a time scale of 1.0--1.7\,h. The refined analysis of the same data  by \citet{marconi2010} revealed  
pulsations at four frequencies with pulsation periods between 0.9 and 2\,h. 
The relatively low projected rotation velocity, $v\,\sin\,i=12\pm2$\,km\,s$^{-1}$, and the fundamental parameters, 
$T_{\rm eff}=6\,500$\,K 
and $\log\,g=3.5$, were recently determined by \citet{cowley2014}
using high-resolution HARPS spectra.
According to their results, PDS\,2 belongs to the class of Herbig~Ae stars
resembling the chemical composition of $\lambda$~Boo stars.

Due to the relative faintness of PDS\,2 ($V=10.8$), the presence of a magnetic field was not previously studied using 
high-resolution polarimetric spectra. In the past, this star was included in two magnetic studies based on low-resolution 
spectropolarimetric observations 
obtained with FORS\,1 (FOcal Reducer low dispersion Spectrograph) mounted at the 8-m Kueyen
telescope of the VLT. While in the first study by \citet{wade2007} no convincing evidence for the presence of 
a magnetic field in PDS\,2 was found,
\citet{Hubrig2009} reported on the detection 
of a mean longitudinal magnetic field $\left<B_{\rm z}\right>=103\pm29$\,G. 
\citet{Bagnulo2012} disputed this detection after re-examining the FORS\,1 data using their 
semi-automatic measurement procedure.

In Sect.~\ref{sect:obs}, we describe the observations and data reduction, and 
in Sect.~\ref{sect:mf_meas} we discuss the methods and results of our magnetic field measurements.
The density distribution of longitudinal magnetic fields in magnetic late Herbig Be and Herbig Ae stars
is presented in Sect.~\ref{sect:mf_compilation}.
Finally, in Sect.~\ref{sect:disc} 
we discuss the significance of the obtained results for the improvement 
of our knowledge  of the role of magnetic fields in intermediate-mass pre-main sequence stars.

\section{Observations and data reduction}
\label{sect:obs}

Two spectropolarimetric observations of PDS\,2 were obtained with the HARPS polarimeter
(HARPSpol; \citealt{snik2008})
attached to ESO's 3.6\,m telescope (La~Silla, Chile) on the nights of 2012 July 15 and 18.
These spectra were originally recorded as part of ESO programme 187.D-0917(C) (PI: Alecian)
and were downloaded from the ESO archive under request MSCHOELLER 77411.
Each observation was split into eight subexposures with an exposure time of 15\,min, 
obtained with different orientations of the quarter-wave retarder plate relative to the 
beam splitter of the circular polarimeter. The time to complete the cycle of eight subexposures for each 
observation accounted for 2\,h and 4\,min.
The achieved signal-to-noise ratio ($S/N$) in the final Stokes~$I$ spectra summed over eight subexposures is rather low, 
accounting for a $S/N=74$ during the night of 2012 July 15, and a $S/N=65$ for the second night on 2012 July 18.
The spectra have a resolving power of about $R = 115\,000$ and
cover the spectral range  3780--6910\,\AA{}, with a small gap between 5259 and 5337\,\AA{}.
The reduction and calibration of these spectra was performed
using the HARPS data reduction software available at the ESO headquarter in Germany.

The normalisation of the spectra to the continuum level consisted of several steps described in detail 
by \citet{Hubrig2013b}.
The Stokes~$I$ and $V$ parameters were derived following the ratio method
\citep{TinbergenRutten1992,Donati1997}, and null polarisation spectra 
were calculated by combining the subexposures 
in such a way that polarisation cancels out.
These steps ensure that no spurious signals are present in the obtained data (e.g.\ \citealt{Ilyin2012}).

\section{Magnetic field measurements}
\label{sect:mf_meas}

\begin{table}
\caption[]{
Magnetic field measurements of the Herbig~Ae star PDS\,2 using the SVD method and the moment technique.
All quoted errors are 1$\sigma$ uncertainties.
}
\label{tab:log_meas}
\centering
\begin{tabular}{cc|cr@{$\pm$}l|r@{$\pm$}lr@{$\pm$}l}
\hline \hline\\[-7pt]
\multicolumn{1}{c}{} &
\multicolumn{1}{c}{} &
\multicolumn{3}{|c}{SVD} &
\multicolumn{4}{|c}{Moment technique} \\
\multicolumn{1}{c}{HJD} &
\multicolumn{1}{c}{S/N} &
\multicolumn{1}{|c}{S/N$_{\rm SVD}$} &
\multicolumn{2}{c}{$\left<B_{\rm z}\right>_{\rm SVD}$} &
\multicolumn{2}{|c}{$\left<B_{\rm z}\right>_{\rm Fe}$} &
\multicolumn{2}{c}{$\left<B_{\rm z}\right>_{\rm Fe,n}$} \\
\multicolumn{1}{c}{} &
\multicolumn{1}{c}{} &
\multicolumn{1}{|c}{} &
\multicolumn{2}{c}{[G]} &
\multicolumn{2}{|c}{[G]} &
\multicolumn{2}{c}{[G]} \\
\hline\\[-7pt]
 2456123.906    & 74 & 2420 &  5 & 5 & 23 & 25 & 22 & 26 \\
 2456126.904    & 65 & 2350 & 33 & 5 & 85 & 26 & 18 & 28 \\
\hline\\
\end{tabular}
\end{table}

The software packages used by our group to study stellar magnetic fields are the moment technique 
developed by \citeauthor{Mathys1991} (e.g.\ \citeyear{Mathys1991}) and the so-called  
multi-line Singular Value Decomposition (SVD) method 
for Stokes profile reconstruction recently introduced by \citet{carroll2012}.
Usually, before we apply the SVD method, we examine the presence of wavelength shifts between right- and left-hand 
side circularly polarized spectra (interpreted in terms of a longitudinal magnetic field $\left<B_{\rm z}\right>$) 
in a sample of clean unblended spectral lines using the  moment technique. 
This technique 
allows us not only the estimation of the mean longitudinal magnetic field, but 
also to prove the presence of crossover effect and of the quadratic magnetic field. This information is 
highly important as, depending on the magnetic field geometry,
even stars with rather weak longitudinal magnetic fields can exhibit strong
crossover effects and kG quadratic fields (see e.g. \citealt{Mathys1995}; \citealt{LandstreetMathys2000}; 
\citealt{MathysHubrig1997,MathysHubrig2006}).
In the study of PDS\,2, for each line in the selected sample of 44 unblended Fe~{\sc ii} lines, the measured shifts 
between the line profiles in the 
left- and right-hand circularly polarized HARPS spectra
are used in a linear regression analysis in the
$\Delta\lambda$ versus $\lambda^2 g_{\rm eff}$ diagram, following the formalism discussed by 
\citet{Mathys1991,Mathys1994}. 
Our measurements $\left<B_{\rm z}\right>=23\pm25$\,G and $\left<B_{\rm z}\right>=82\pm26$\,G achieved on 
the first and second epochs, respectively, indicated the probable presence of a weak magnetic field
on the second epoch. No crossover or mean quadratic magnetic field at a significance level of 3$\sigma$ has been 
detected on either observing night, probably due to the rather low S/N of the observed spectra. 
The measured values for the mean longitudinal magnetic 
field $\left< B_{\rm z} \right>$ are presented in the fifth column of Table~\ref{tab:log_meas}. 
In the last column, we also present the results of the magnetic field measurements 
using the null polarisation spectra, labeled with $n$.
Since no significant fields could be 
determined from the null spectra, we conclude that no noticeable spurious polarisation is present. 

The basic idea of the SVD method is similar to the Principal Component Analysis (PCA) approach
\citep{MG2008,Semel2009}, where 
the similarity of the individual Stokes~$V$ profiles allows one to describe the most coherent and 
systematic features present in all spectral line profiles as a projection onto a small number of 
eigenprofiles.

\begin{figure}
\centering
\includegraphics[width=0.22\textwidth]{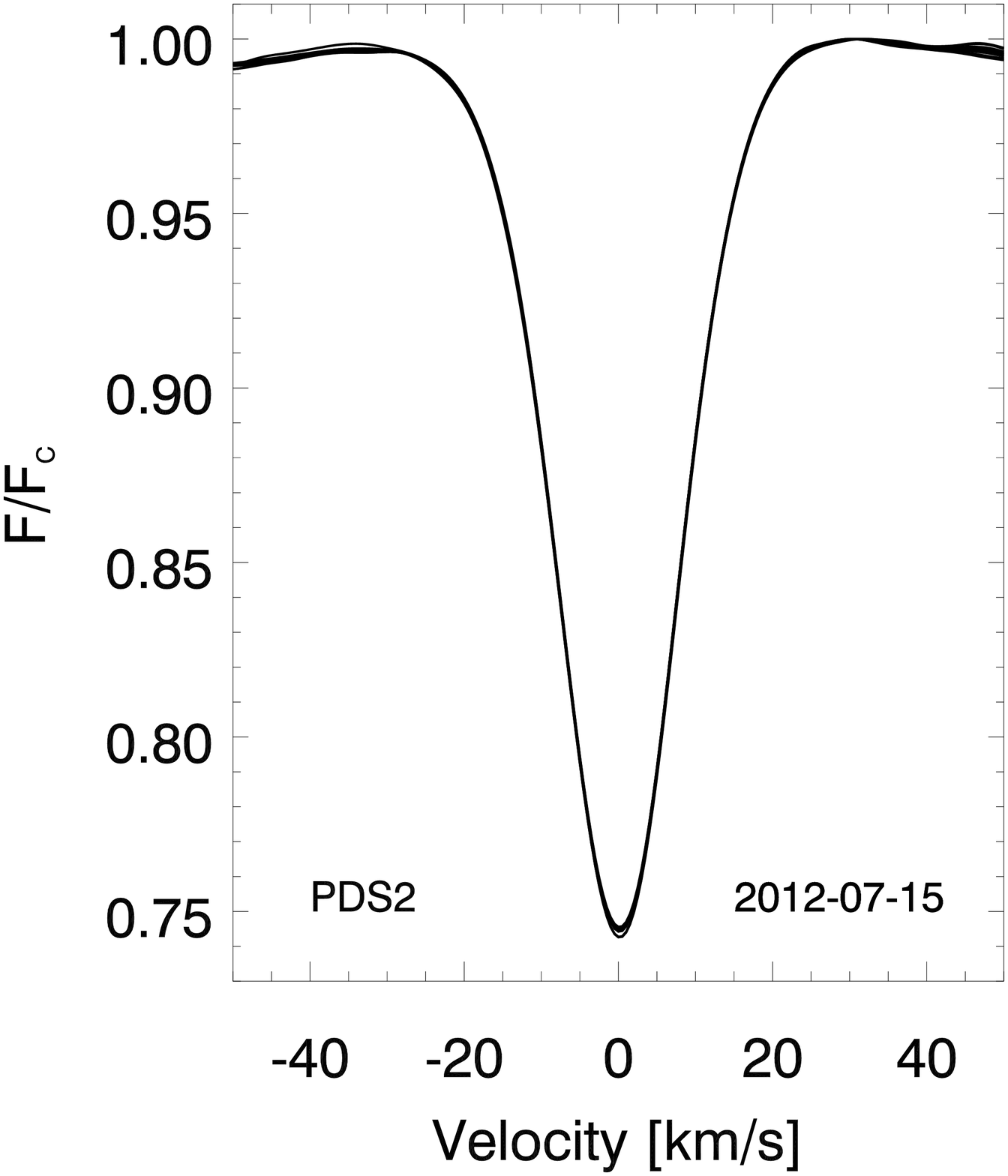}
\includegraphics[width=0.22\textwidth]{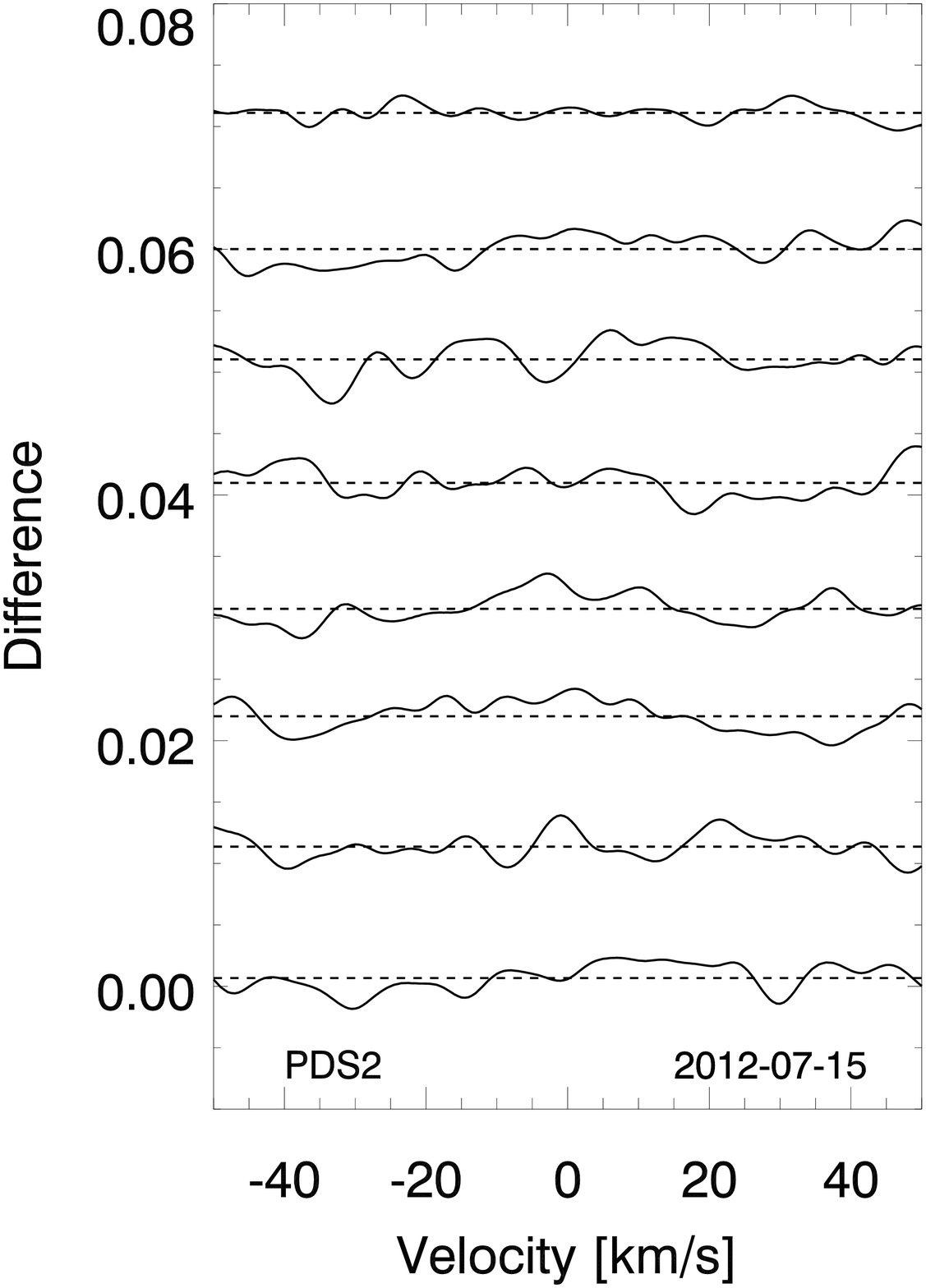}
\includegraphics[width=0.22\textwidth]{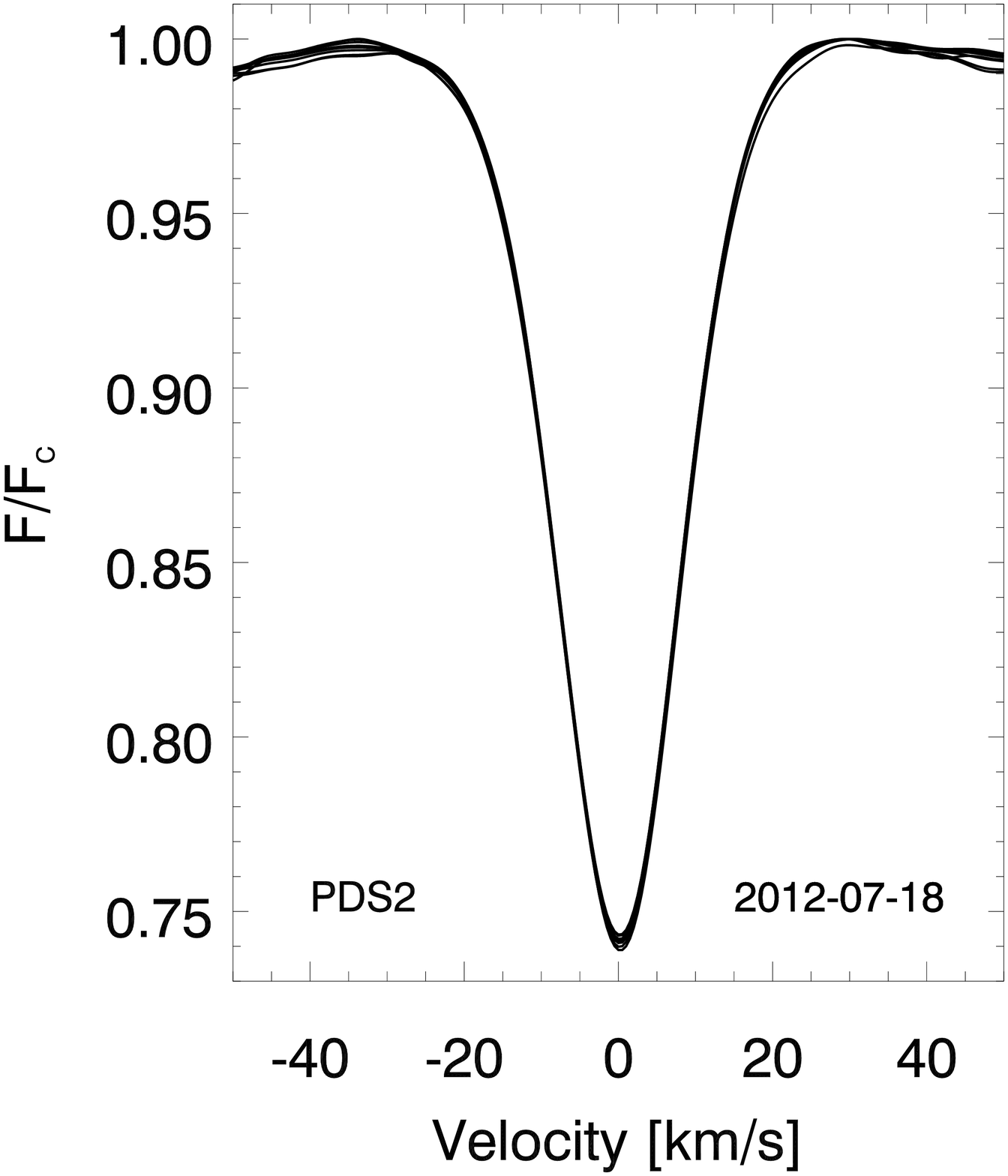}
\includegraphics[width=0.22\textwidth]{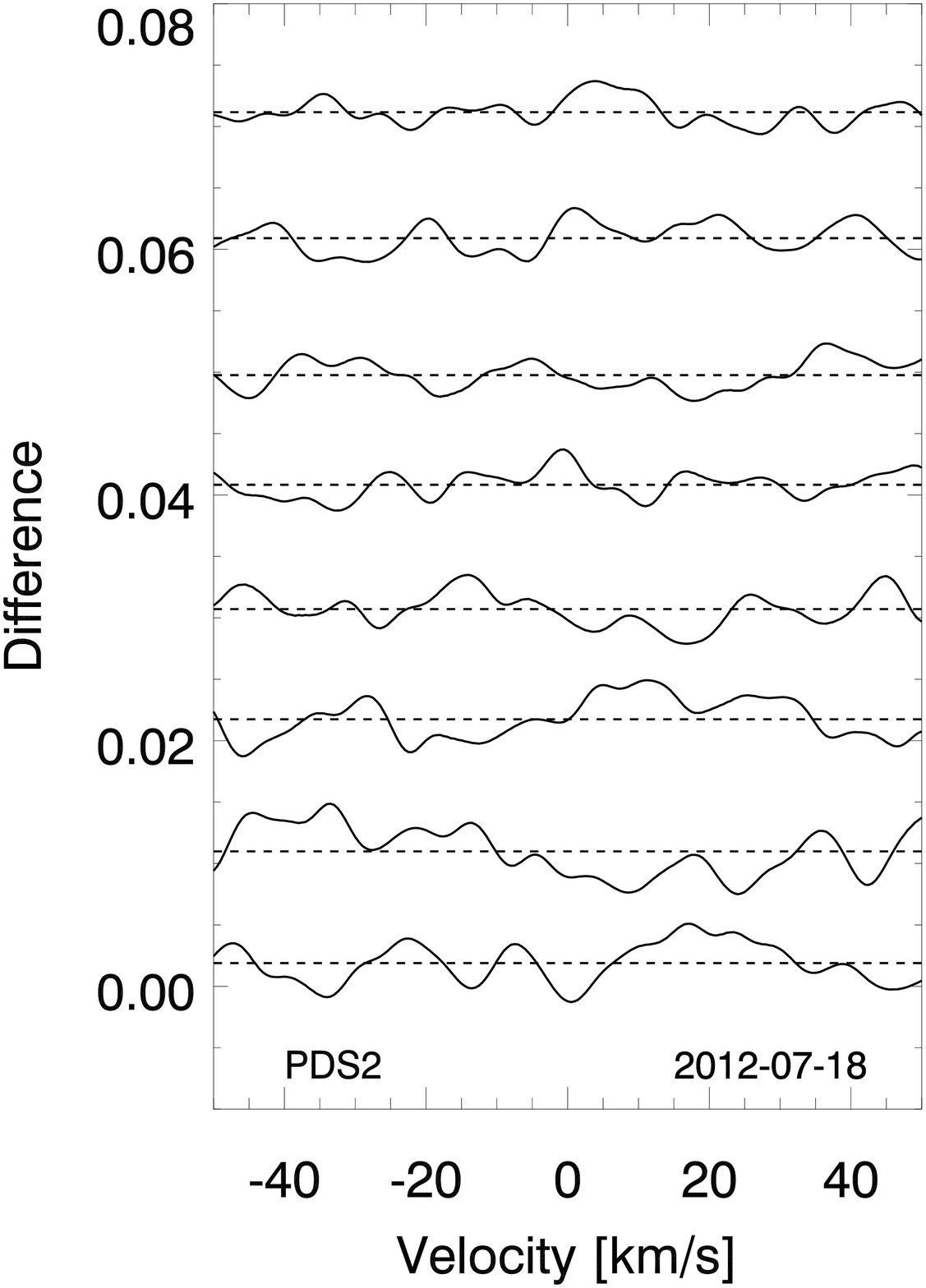}
\caption{Comparison of the SVD Stokes~$I$ profiles in the subexposures recorded on the nights of 2012 July 15 (top)
and 18 (bottom).  
{\it Left panels}:
Overplotted Stokes~$I$ profiles computed for the individual subexposures obtained with a time lapse of 15\,min. 
{\it Right panels}:
Differences between the Stokes~$I$ profiles computed for the individual subexposures and the average Stokes~$I$ profile.
}
\label{fig:night12}
\end{figure}

As we mentioned above, the star PDS\,2 was reported to show $\delta$~Scuti-like pulsations on a time scale
between 0.9 and 2.0\,h. Since the full HARPS exposure times of about 2\,h on both nights are 
of the same order, and pulsations are known to have an impact 
on the analysis of the presence of a magnetic field and its strength (e.g., \citealt{schnerr2006};
\citealt{Hubrig2011b}), as a first step, we verified that no change 
in the line profile shape or radial velocity shifts are present in the obtained spectra.
In Fig.~\ref{fig:night12} on the left side we present overplotted Stokes~$I$ profiles computed for the individual 
subexposures recorded on both nights. On the right side of this figure we show differences between the Stokes~$I$ profiles 
computed for the individual subexposures and the average Stokes~$I$ profile.
No impact of pulsations at a level higher than the spectral noise is detected in the Stokes~$I$ profiles.
 
\begin{figure}
\centering
\includegraphics[width=0.45\textwidth]{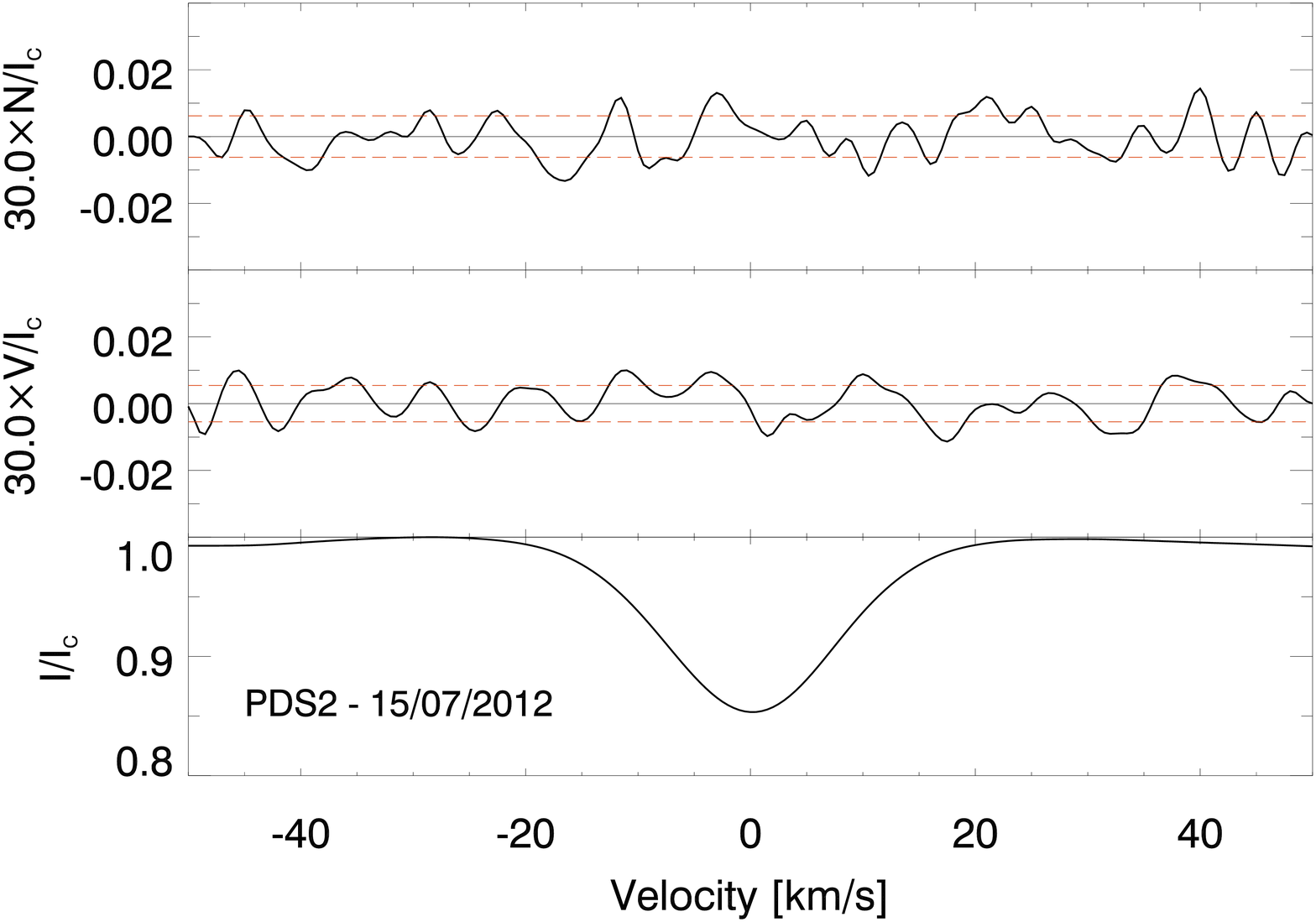}
\includegraphics[width=0.45\textwidth]{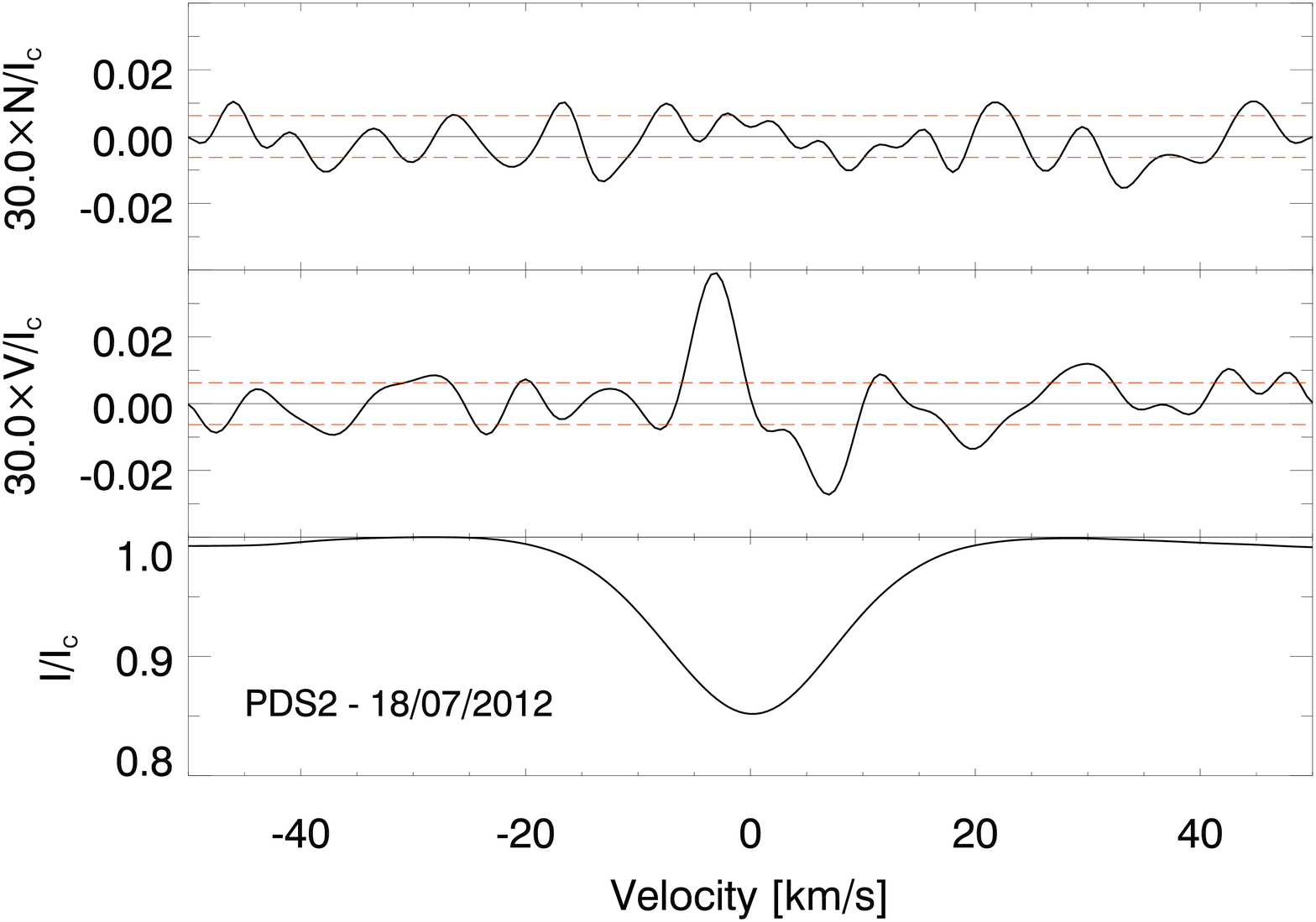}
\caption{
$I$, $V$, and $N$ SVD profiles  obtained for PDS\,2 on two different nights.
The $V$ and $N$ profiles were expanded by 
a factor of 30 and shifted upwards for better visibility. The red (in the on-line version) dashed lines
indicate the standard deviations for the $V$ and $N$ spectra.}
\label{fig:SVD}
\end{figure}

The resulting mean Stokes~$I$, Stokes~$V$, and Null profiles obtained by using the SVD
method with 2094 metallic lines in the line mask are presented in Fig.~\ref{fig:SVD}.
The line mask was constructed using the VALD database 
(e.g.\ \citealt{kupka2000}) and the respective stellar parameters of PDS\,2.
Due to the very low signal-to-noise at the blue and red ends of the HARPS spectra, the spectral lines
used for the computation of the SVD profiles have been selected in the wavelength 
region from 4200 to 6200\,\AA{}.
The mean longitudinal magnetic field is estimated from the SVD reconstructed Stokes~$V$ and $I$
using the center-of-gravity method (see e.g.\ \citealt{CarrollStrassmeier2014}). A velocity range
$\pm 21$\,km\,s$^{-1}$ was adopted to determine the detection probability and the longitudinal magnetic field value.
For both observations the null spectra appear flat, indicating the absence of spurious polarisation.
No longitudinal magnetic field is detected on the first epoch on 2012 July 15, where we measure 
$\left<B_{\rm z}\right>=5\pm5$\,G.
A definite detection of a field, $\left<B_{\rm z}\right>=33\pm5$\,G, with
a false alarm probability (FAP) smaller than $10^{-6}$,
is achieved on the second epoch three nights later. 

In Table~\ref{tab:log_meas}.
we list for the two observations the heliocentric Julian date, the $S/N$ reached in the final Stokes~$I$ profiles,
the $S/N$ obtained in the SVD profile, the longitudinal magnetic field $\left<B_{\rm z}\right>$ determined with SVD,
and $\left<B_{\rm z}\right>$ determined with the moment technique from iron lines.

\section{Magnetic field measurements reported in previous spectropolarimetric studies}
\label{sect:mf_compilation}

As the magnetic field 
strength and the magnetic field geometry in Herbig stars are poorly known -- only about 20 late Herbig Be and Herbig Ae 
stars have been reported to possess large-scale organized magnetic fields --  we compiled
for them all magnetic field measurements reported in previous spectropolarimetric studies.
In Table~\ref{tab:4} we present the $\overline{\left< B_{\rm z} \right>}$ values 
selected from previous low-resolution spectropolarimetric studies with FORS\,1/2, where
magnetic field measurements were carried out using the entire spectra,
and high-resolution spectropolarimetric studies using the HARPS, ESPaDOnS, and Narval spectrographs.
The rms longitudinal magnetic field, the rms standard error,  and the reduced $\chi^2$ for 
these measurements have been computed following the equations of \citet{bo83}:

\begin{eqnarray}
\overline{\left< B_{\rm z} \right>} &=& \left( \frac{1}{n} \sum^{n}_{i=1} \left< B_{\rm z} \right> ^2_i \right)^{1/2}, \nonumber \\
\overline{\sigma} &=& \left( \frac{1}{n} \sum^{n}_{i=1} \sigma^2_i \right)^{1/2}, \nonumber \\
\chi^2/n &=& \frac{1}{n} \sum_{i=1}^n \left( \frac{\left< B_{\rm z} \right>_i}{\sigma_i} \right)^2. \nonumber
\end{eqnarray}

Table~\ref{tab:4} lists for each star its name and spectral type, the number of magnetic field
measurements used to calculate $\overline{\left< B_{\rm z} \right>}$, the rms longitudinal magnetic
field $\overline{\left< B_{\rm z} \right>}$, the rms standard error,
the reduced $\chi^2$ values, and finally the references to the measurements.

\begin{table*}
\caption{
rms longitudinal magnetic field strength, rms standard error, and reduced $\chi^2$ values of late Herbig~Be and Herbig~Ae stars.
$N$ gives the number of measurements for the individual targets, separately for the low resolution
spectrographs FORS\,1 and FORS\,2 and the high resolution spectrographs HARPS, ESPaDOnS, and Narval.
}
\label{tab:4}
\begin{center}
\begin{tabular}{lcrrrrrl}
\hline
\hline
\multicolumn{1}{c}{\raisebox{2mm}{\rule{0mm}{2mm}}Name} &
\multicolumn{1}{c}{Sp.~T.} &
\multicolumn{1}{c}{$N_{\rm low res}$} &
\multicolumn{1}{c}{$N_{\rm hi res}$} &
\multicolumn{1}{c}{$\overline{\left< B_{\rm z} \right>}$} &
\multicolumn{1}{c}{$\overline{\sigma}$} &
\multicolumn{1}{c}{$\chi^2/n$} &
\multicolumn{1}{c}{References}\\
\multicolumn{1}{c}{} &
\multicolumn{1}{c}{} &
\multicolumn{1}{c}{} &
\multicolumn{1}{c}{} &
\multicolumn{1}{c}{[G]} &
\multicolumn{1}{c}{[G]} &
\multicolumn{1}{c}{} &
\multicolumn{1}{c}{}\\
\hline
PDS\,2      & F2 &  3 &2 &   75 &  25 & 14.69 & W07 H09 H15 \\
HD\,31648   & A3 &  2 &5 &  416 & 125 &  9.25 & H07 W07 H11b A13a \\
HD\,35929   & F2 &  1 &5 &   54 &  23 &  5.84 & W07 A13a \\
HD\,36112   & A5 &  1 &2 &   89 &  84 &  1.11 & W07 A13a \\
V380\,Ori   & A1 &  3 &24 &  348 & 137 & 10.09 & W05 W07 A09 \\
BF\,Ori     & A2 &  1 &2 &   87 &  36 & 16.18 & W07 A13a \\
HD\,58647   & B9 &  0 &1 &  218 &  69 &  9.98 & H13 \\
Z\,CMa      & B9 &  1 &0 & 1231 & 164 & 56.34 & S10 \\
HD\,97048   & A0 & 19 &0 &  105 &  58 &  4.64 & W07 H09 H11b \\
HD\,98922   & A2 &  1 &2 &  135 &  64 &  6.33 & W07 A13a H13 \\
HD\,100546  & B9 &  2 &0 &  106 &  52 &  7.40 & W07 H09 \\
HD\,101412  & A0 & 16 &0 &  273 &  53 & 33.11 & W05 W07 H09 H11a \\
HD\,104237  & A4 &  3 &2 &   56 &  35 &  5.75 & D97 W07 H13 \\
HD\,135344A & A0 &  2 &1 &   80 &  85 &  5.76 & H09 A13a \\
HD\,139614  & A7 &  6 &3 &   73 &  26 &  8.33 & W05 H07 H09 A13a \\
HD\,144432  & A7 &  6 &1 &  100 &  50 &  3.52 & H07 W07 H09 A13a \\
HD\,144668  & A7 &  2 & 3&  106 &  34 &  8.55 & H07 W07 H09 A13a \\
HD\,150193  & A1 & 15 &1 &  159 & 136 &  6.84 & H09 H11b A13a \\
HD\,176386  & B9 & 15 &1 &  130 &  81 &  4.28 & H09 H11b A13a \\
HD\,190073  & A1 & 6 &68 &   62 &  21 & 16.10 & C07 H07 W07 H09 A13b H13 \\
\hline
\end{tabular}
\begin{flushleft}
References:
A09 -- \citet{alecian2009};
A13a -- \citet{alecian2013a};
A13b -- \citet{alecian2013b};
C07 -- \citet{Catala2007};
D97 -- \citet{Donati1997};
H07 -- \citet{Hubrig2007};
H09 -- \citet{Hubrig2009};
H11a -- \citet{Hubrig2011a};
H11b -- \citet{Hubrig2011b};
H13 -- \citet{Hubrig2013b};
H15 -- this paper;
S10 -- \citet{Szeifert2010};
W05 -- \citet{Wade2005};
W07 -- \citet{wade2007}.
\end{flushleft}
\end{center}
\end{table*}

\begin{figure}
\centering
\includegraphics[width=0.35\textwidth]{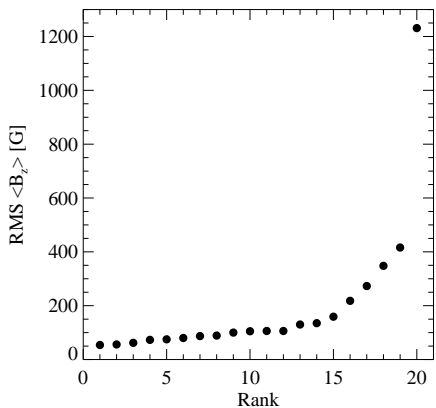}
\caption{
Density distribution of the rms longitudinal magnetic field values (Col.~4 in Table~\ref{tab:4})
for the twenty late Herbig~Be and Herbig~Ae stars
for which detections of a magnetic field were reported in the past.
}
\label{fig:rms}
\end{figure}

In Fig.~\ref{fig:rms}, we present the distribution of the rms longitudinal magnetic field strength
for all late Herbig~Be and Herbig~Ae stars, for which detections of a magnetic field were reported in the past.
The obtained density distribution of the rms longitudinal magnetic field values 
reveals that only very few stars have 
rms fields stronger than 200\,G, and half of the sample possesses magnetic fields
of about 100\,G and less.

\section{Discussion}
\label{sect:disc}

In this work, we present the first detection of a weak magnetic field in the Herbig~Ae star
PDS\,2 using high-resolution spectropolarimetric observations with HARPS. Previous observations
using low-resolution spectropolarimetry with FORS\,1 indicated  $\left<B_{\rm z}\right>=103\pm29$\,G \citep{Hubrig2009}.
Apart from PDS\,2, the authors 
announced the detection of a weak longitudinal magnetic fields in six other Herbig~Ae stars.
Confirmation of the definite presence of the magnetic field
based on accurate high-resolution spectropolarimetry with low uncertainties
is still pending for these other six stars.
Among these Herbig~Ae/Be stars, three show longitudinal magnetic 
fields below 100\,G, while for the remaining three stars the magnetic field ranges from 100 to 200\,G. 

Magnetospheric accretion has been well established for T\,Tauri stars and depends on a strong ordered 
predominantly dipolar field channeling circumstellar disk material to the stellar surface via accretion streams.
However, no convincing scenario on how magnetospheric accretion works in Herbig~Ae stars possessing rather 
weak magnetic fields currently exists. 
\citet{caul2014} recently studied He~{\sc i} $\lambda$10830 morphology in a sample of 
56~Herbig~Ae/Be stars. They suggest that Herbig~Be stars do not accrete material from their inner
disks in the same manner as T\,Tauri stars, while late Herbig~Be and Herbig~Ae stars show evidence for 
magnetospheric accretion.
Further, due to their high rotation rates \citep{Muzerolle2004} and weak 
magnetic fields, more compact magnetospheres in Herbig stars are proposed.
We note that since the rotation period of PDS\,2 is currently unknown, it is not clear
whether it might be a rapid rotator with a low inclination of the rotation axis.

In fact, the density distribution of the rms longitudinal magnetic field strength clearly indicates that 
the magnetic fields in Herbig~Ae stars are by far weaker than those measured in their lower mass
counterparts, the T\,Tauri stars, usually possessing kG magnetic fields.
As is shown in Fig.~\ref{fig:rms}, out of the sample of twenty magnetic late Herbig~Be and Herbig~Ae stars, 
only very few stars have rms fields stronger than 200\,G, and half of the sample possesses magnetic fields
of about 100\,G and less. 
The obtained results seem to support the existence of a different accretion mechanism 
mediating the mass flow from the disk to the intermediate-mass star, compared to that working in T\,Tauri stars.
Noteworthy, for the currently best studied Herbig~Ae stars, HD\,101412 
and V380\,Ori with strong magnetic fields monitored over the rotation cycles, the obtained magnetic 
field models assuming a simple centered dipole indicate rather large obliquity $\beta$ of 
the magnetic axis to the rotation axis \citep{Hubrig2011a,alecian2009}.
The fact that the dipole axes are located close to the stellar equatorial 
plane is very intriguing in view of the generally assumed magnetospheric accretion scenario.
As was shown in the past \citep{Romanova2003},
the topology of the channeled accretion critically depends on the 
tilt angle between the rotation and the magnetic axis. For large inclination angles $\beta$,
many polar field lines would thread the inner region of the disk, while the closed 
lines cross the path of the disk matter, causing strong magnetic braking. 
Clearly, the qualitative picture is expected to be different if the magnetic field topology was proven as more complex than a 
simple centered dipole model.

For a better understanding of the role of magnetic fields in 
Herbig~Ae/Be stars, 
it is important to carry out additional highly accurate magnetic field measurements of a representative 
sample of Herbig~Ae stars 
over their rotation periods using high quality, 
high-resolution polarimetric spectra.
Using multi-epoch observations, it will become possible to disclose the magnetic topology on the surface 
of Herbig~Ae stars necessary to understand the interaction of the magnetic field with winds, accretion disks, 
convection, turbulence, and circulation.

\section*{Acknowledgments}
\label{sect:ackn}

Based on data obtained from the ESO Science Archive Facility under request MSCHOELLER 77411.
Based on observations made with ESO Telescopes at the La Silla Paranal Observatory under programme ID 187.D-0917(C).
We would like to thank the anonymous referee for useful comments.

\label{lastpage}

\end{document}